\documentclass[12pt]{article}
\usepackage{epsfig,fullpage}

\begin{document}
\begin{center}

\begin{flushright}
     SWAT/346 \\
August 2002
\end{flushright}
\par \vskip 10mm

\vskip 1.2in

\begin{center}
{\LARGE\bf
Non-Compact QED$_3$ with $N_f \geq 2$} 

\vskip 0.7in
S.J. Hands $\!^a$, J.B. Kogut $\! ^b$ and
 C.G. Strouthos $\!^{a}$ \\
\vskip 0.2in

$^a\,${\it Department of Physics, University of Wales Swansea,\\
Singleton Park, Swansea, SA2 8PP, U.K.} \\
$^b\,${\it Department of Physics, University of Illinois at Urbana-Champaign,\\
Urbana, Illinois 61801-3080, U.S.A.}\\
\end{center}

\vskip 1.0in 
{\large\bf Abstract}
\end{center}
\noindent
Non-compact three-dimensional QED is studied by computer simulations to understand
its chiral symmetry breaking features for $N_f \geq 2$, on lattice volumes
up to $50^3$ and bare masses as low as $m_0a=0.0000625$.
We compute the
chiral condensate, scalar and pseudoscalar susceptibilities, and the
masses of scalar and pseudoscalar mesons. Finite volume effects 
and discretisation artifacts are carefully monitored.
Our results reveal no 
decisive signal for chiral symmetry breaking for any $N_f\geq2$. 
For $N_f=2$ the dimensionless condensate can be bounded by
$\beta^2 \langle \bar\Psi \Psi \rangle \leq 5 \times 10^{-5}$.
We also present an exploratory study of the fractionally-charged Polyakov line.

\newpage

\section{Introduction}

Over the past few years QED$_3$ has attracted a lot of attention, because of
potential applications to models of high $T_c$ superconductivity \cite{dorey}.
It is also an interesting  and challenging model field theory in its own right,
which is
seen as an ideal laboratory to study effects such as dynamical mass
generation which may be relevant to phenomena described by
more complicated gauge field theories such as technicolor
\cite{walking}.
In three dimensions the 
model is super-renormalisable, because 
the coupling $g$ has mass dimension ${1\over2}$ and so provides the 
intrinsic scale of the theory independently of any bare fermion mass $m_0$. 
This also implies asymptotic freedom, 
since for processes with momentum transfer $k\gg g^2$
the theory is effectively non-interacting. 

Non-trivial behaviour may arise in
the infra-red, however, as suggested by an expansion in $1/N_f$, where $N_f$ is
the number of (four-component) fermion species \cite{pisarski}. 
In the large-$N_f$ limit
the photon propagator 
is modified by the vacuum polarisation diagram 
from $1/k^2$ to $1/[k^2+{g^2\over8}N_fk]$; in coordinate space the logarithmic 
Coulomb potential is thus modifed to $1/r$ for distances $r\gg (g^2N_f)^{-1}$.
Naively we deduce that the confining property of the Coulomb potential
is screened by virtual $f\bar f$ pairs. However, consider the energy required to
excite a real $f\bar f$ pair into the vacuum state \cite{sasha}. 
The kinetic energy of the pair is positive and scales as $r^{-1}$ 
by the uncertainty
principle.
For large $r$ therefore, both kinetic and potential terms 
scale as $r^{-1}$ and it becomes a delicate question which dominates. Next we
note that the dimensionless
interaction strength scales in this regime as $N_f^{-1}$, implying
the possible existence of a critical $N_{fc}$ below which the negative
potential term
dominates as $r\to\infty$. Since as $r\to0$ the positive kinetic term must
dominate the logarithmic Coulomb term, we deduce in this case the existence of 
an energy  minimum at some non-zero $r$, 
implying the existence of stable $f\bar f$ bound states in the ground
state. This semi-classical argument hence suggests a non-vanishing
condensate $\langle\bar\Psi\Psi\rangle\not=0$ for $N_f<N_{fc}$. Since the 
condensate spontaneously breaks chiral symmetry, 
the fermions are imbued with a
dynamically-generated mass. For asymptotically large $r$ 
the massive fermions decouple, ceasing to
screen the charge, and the
logarithmically-confining Coulomb potential is restored. 
The picture therefore predicts two possible phases for QED$_3$:
for $N_f<N_{fc}$ chiral symmetry is broken, so that the only massless states are
Goldstone bosons, and electric charge is logarithmically confined; for
$N_f>N_{fc}$ the theory is conformal, 
consisting of massless fermions interacting
via a long-range $1/r$ interaction.

In this paper we address the question of 
whether chiral symmetry is spontaneously
broken in QED$_3$ for any value of $N_f \geq 2$, i.e. 
whether $\langle \bar{\Psi} \Psi \rangle \neq 0$ 
in the chiral limit bare fermion mass $m_0 \rightarrow 0$ 
and the continuum (weak coupling)
limit $g \rightarrow 0$. This question has received much study over the 
past decade, the principal approach being self-consistent solutions to
truncated Schwinger-Dyson (SD) equation for the fermion propagator.
Initial studies 
using the
photon propagator derived from the leading order large-$N_f$ expansion
suggested that
the answer is positive
with $N_{fc}\simeq3.2$ \cite{Nfc}. 
Within this framework it was shown that the dynamically generated fermion 
mass is much less than the coupling contant, providing a natural
hierarchy of mass scales. It is also possible to consider analytically
continuing to a continuous number of flavors;
the model in the limit
$N_f\to N_{fc}$ is then supposed to undergo an infinite-order phase transition
\cite{miranskii}. 
Other studies
taking non-trivial vertex corrections into account, however,
predicted chiral symmetry
breaking for arbitrary $N_f$ \cite{pennington}. 
More recent studies which treat
the vertex consistently in both numerator and denominator of the SD equations
have found $N_{fc}<\infty$, with a value either in agreement with the
original study
\cite{maris}, or a slightly higher $N_{fc}\simeq4.3$ \cite{nash}.
Other analytical approaches have also been applied.
A recent renormalization group analysis
predicts $3<N_{fc}<4$ \cite{kubota}.
Finally there is an argument based on the inequality 
\begin{equation}
f_{IR}\leq f_{UV},\label{eq:appel}
\end{equation}
where $f$ is the absolute value of the thermodynamic free
energy, and can be estimated in the relevant limits simply by counting
massless degrees of freedom, either fermions or Goldstones as reviewed below.
For QED$_3$ this yields the prediction
$N_{fc}\leq{3\over2}$ \cite{appelquist}. 

There have also been attempts to resolve the issue via numerical
simulation of
non-compact lattice QED$_3$. 
The obvious advantage of a computer simulation is that 
one can study any $N_f$ without any technical assumption 
concerning the convergence of 
expansion methods. 
Indeed, it is also possible to study the quenched ($N_f=0$) limit, in which case
a numerical study has shown that chiral symmetry appears to be broken
\cite{hands}, with a 
dimensionless condensate
$\beta^2\langle\bar\Psi\Psi\rangle\simeq5\times10^{-3}$, where
$\beta\equiv1/g^2a$ is the dimensionless lattice coupling constant
($a$ being the
lattice spacing).
For the unquenched theory, opinions have divided
on whether $N_{fc}$ is
finite and $\simeq3$ \cite{sasha,dagotto}, 
or whether chiral symmetry is broken for all
$N_f$ \cite{azcoiti}. The most recent study with $N_f=2$ on lattices up to 
$16^3$ found chiral symmetry breaking with
$\beta^2\langle\bar\Psi\Psi\rangle\simeq3\times10^{-3}$ \cite{alex}.
This value is extracted, however, by simultaneous extrapolation to 
continuum, chiral, and thermodynamic limits, which is difficult 
to control since even with broken chiral symmetry there remain
large finite volume effects due to the presence of a massless photon in
the spectrum. For this reason it is unlikely that a numerical simulation will
ever completely exclude chiral symmetry breaking for $N_f>0$, 
particularly if the 
condensate is suppressed by a typical non-perturbative factor $\exp(-CN_f)$
\cite{pennington}.

In this paper we will present results of simulations on lattices 
considerably larger (up to $50^3$) and bare fermion masses
considerably smaller ($m_0a\geq0.0000625$) 
than any previous study, which enables us
to approach all three limits with greater control. Our goal is to place a bound
on any possible chiral condensate using as few assumptions as possible on the
nature of the required extrapolations. Our conclusion is 
that for $N_f=2$, $\beta^2\langle\bar\Psi\Psi\rangle<5\times10^{-5}$,
two orders of magnitude below the quenched value. For larger $N_f$ the
condensate is smaller still. Since we see no direct evidence for chiral
symmetry breaking, our results are consistent with $N_{fc}<2$, although as
explained above an exponentially suppressed condensate is almost impossible to
exclude. In the next section we briefly review the model, its symmetries, and
the numerical methods used to simulate it. Our numerical results for the 
condensate
appear in Section 3, in which in addition we present results for meson masses
and susceptibilities, and Polyakov lines with both integer and fractional
charges. All of these subsidiary quantities reveal evidence of strong finite
volume effects which makes any conclusions drawn qualitative at best.
Section 4 summarises our findings.

\newpage
\section{The Model}
The 
continuum Lagrangian density describing QED$_3$ is given in Euclidean
metric by
\begin{equation}
{\cal L} = \frac{1}{4}F_{\mu\nu}F_{\mu\nu}
+ {\bar \Psi}_i D_\mu\gamma _\mu \Psi_i +m_0 {\bar
 \Psi}_i \Psi_i
\label{eq:contmodel}
\end{equation}
where $D_\mu \equiv \partial _\mu -ig A_\mu $,
and $F_{\mu\nu}$ is the field
strength for the abelian gauge field $A_\mu$.
The fermions $\Psi_i$, $i=1,\ldots,N_f$,
are four-component spinors acted on by hermitian Dirac matrices $\gamma_\mu$, 
implying that the mass term proportional to $m_0$ is
invariant under the 2+1 dimensional parity transformation.

Since the $\gamma$-matrices belong to a reducible representation of the Dirac
algebra in three dimensions, the global symmetry of (\ref{eq:contmodel}) is
larger than naively expected.
The matrices
$\{ {\bf 1}, \gamma _3, \gamma _5,
\tau_3\}$
where $\tau_3 \equiv i \gamma_3 \gamma_5$ 
generate a global U(2) symmetry.
To see this define $\tilde\Psi\equiv\bar\Psi\tau_3$.
For $m_0=0$ the U(2) symmetry is then
\begin{equation}
\Psi\rightarrow U\Psi\qquad\tilde\Psi\rightarrow\tilde\Psi U^\dagger\qquad
U\in\mbox{U(2)}.
\end{equation}
The symmetry is broken, either explicitly by a bare mass or spontaneously 
by chiral condensation $\langle\bar\Psi\Psi\rangle
\not=0$, to a residual U(1)$\otimes$U(1)$_{\tau_3}$
where the first U(1) factor corresponds to global fermion number conservation,
and the second to a symmetry
\begin{equation}
\Psi\rightarrow e^{i\alpha\tau_3}\Psi\qquad
\bar\Psi\rightarrow\bar\Psi e^{-i\alpha\tau_3}.
\end{equation}
Chiral symmetry breaking for the theory with $N_f$ flavors 
would then result in the pattern
$\mbox{U}(2N_f)\rightarrow\mbox{U}(N_f)\otimes\mbox{U}(N_f)_{\tau_3}$.

The lattice action using staggered lattice fermion fields
$\chi,\bar\chi$
is given by 
\begin{eqnarray}
S &=&\frac{\beta}{2} \sum_{x,\mu<\nu} F_{\mu \nu}(x) F_{\mu \nu}(x)
+ \sum_{i=1}^N \sum_{x,x^\prime} {\bar \chi}_i(x) M(x,x^\prime) 
\chi_i(x^\prime)
\label{eq:action}\\
F_{\mu \nu}(x) &\equiv& \theta_{x\mu}+\theta_{x+\hat\mu,\nu}
-\theta_{x+\hat\nu,\mu}-\theta_{x\nu}\nonumber\\
M(x,x^\prime) &\equiv&  
m_0 \delta_{x,x^\prime}+\frac{1}{2} \sum_\mu\eta_{\mu}(x)
[\delta_{x^\prime,x+\hat \mu} U_{x\mu}
-\delta_{x^\prime,x-\hat \mu} U_{x-\hat \mu,\mu}^\dagger].\nonumber
\end{eqnarray}

The vectors $x,~x^\prime$ consist of three integers
$(x_0,x_1,x_2)$
labelling the sites of an $L^3$ lattice.
Since the gauge action $F^2$ is unbounded from above,
(\ref{eq:action})
defines the {\sl non-compact\/} formulation of lattice QED$_3$.
The $\eta_\mu(x)$ are Kawamoto-Smit phases 
($\eta_1(x)=1;\;\eta_2(x)=(-1)^{x_1};\;\eta_0(x)=(-1)^{x_1+x_2}$)
designed to ensure relativistic covariance in the continuum limit.
For the fermion fields antiperiodic boundary conditions are used in
the timelike direction and periodic
boundary conditions in the spatial directions.
The phase factors in the fermion bilinear are defined by
$U_{x\mu} \equiv
\exp(i\theta_{x \mu})$,
where
$\theta_{x \mu}$ is the gauge
potential.
In terms of continuum quantities,
$\theta_{x\mu}= a gA_\mu(x)$ and the coupling
$\beta \equiv \frac{1}{g^2 a}$,
where $a$ is the physical lattice spacing.
We prefer the non-compact version of lattice QED$_3$, because it resembles
more closely the continuum model we wish to compare with.
In particular, magnetic
monopole excitations, which are known to have a decisive effect on the infra-red
behaviour of the compact model \cite{mono}, 
are here strongly suppressed as $\beta$ is made
large.

The particle content of eq.(\ref{eq:action}) has been analysed
in \cite{burden},
showing that in the long-wavelength limit $N$ 
flavors of staggered lattice fermion correspond to
$N_f=2N$ of the four-component continuum flavors described by the action
(\ref{eq:contmodel}).
For 
lattice spacing $a>0$, however, the global symmetries for $m_0=0$
are only partially
realised. In this case the symmetry is
\begin{eqnarray}
\bar\chi_o\rightarrow\bar\chi_o e^{i\alpha}&~&\qquad
\chi_e\rightarrow e^{-i\alpha}\chi_e \nonumber \\
\bar\chi_e\rightarrow\bar\chi_e e^{i\beta}&~&\qquad
\chi_o\rightarrow e^{-i\beta}\chi_o,
\label{eq:latsym}
\end{eqnarray}
where $\chi_{o/e}$ denotes the field on odd (i.e. $
\varepsilon(x)\equiv(-1)^{x_1+x_2+x_3}=-1$)
and even sublattices respectively. Chiral symmetry breaking characterised by
$\langle\bar\Psi\Psi\rangle\equiv\langle\bar\chi\chi\rangle\not=0$ 
now has the pattern
$\mbox{U}(N)\otimes\mbox{U}(N)\rightarrow\mbox{U}(N)$.
Only in the continuum limit
$\beta\to\infty$ is the full symmetry of the continuum theory expected to be
restored.

In the context of the current discussion the issue of whether the full
U($2N_f$)=U($4N$) flavor symmetry is restored 
in the continuum limit $\beta\to\infty$
is very important. Take for instance the considerations of ref.
\cite{appelquist} which constrain $N_{fc}$ via the inequality (\ref{eq:appel}):
\begin{equation}
1+\textstyle{3\over4}(4N_{fc})\leq1+2N_{fc}^2.\label{eq:thermo1}
\end{equation}
Here the left hand side of the inequality counts the number of massless 
thermodynamic degrees 
of freedom in the UV limit, which consist of a single bose (the photon) and 
$4N_f$ fermi degrees of freedom, the factor ${3\over4}$ coming from 
the Fermi-Dirac
distribution in 2+1 dimensions. The right hand side counts both the photon
and the $2N_f^2$ Goldstone modes expected from the breaking
$\mbox{U}(2N_f)\rightarrow\mbox{U}(N_f)\otimes\mbox{U}(N_f)$. Eq.
(\ref{eq:thermo1}) yields the bound $N_{fc}\leq{3\over2}$. If, however, 
chiral symmetry broke according to the staggered pattern
$\mbox{U}(N)\otimes\mbox{U}(N)\rightarrow\mbox{U}(N)$, then the inequality
becomes
\begin{equation}
1+\textstyle{3\over4}(8N_{c})\leq1+N_{c}^2,\label{eq:thermo2}
\end{equation}
implying $N_{fc}=2N_c\leq12$ \cite{alex}. An observation of broken
chiral symmetry in the $\beta\to\infty$ limit
in lattice QED$_3$ simulations with $N_f\geq2$ therefore implies
either that the arguments leading to the constraint (\ref{eq:appel}) are 
incorrect, or that flavor symmetry restoration does not occur in this model.

The numerical results presented here were obtained by simulating the
action (\ref{eq:action}) using a standard Hybrid Monte Carlo (HMC) algorithm.
The form of (\ref{eq:action}) permits an even-odd partitioning,
which means that a single flavor of staggered
fermion can be simulated. The minimum number of continuum flavors which can be
simulated with a local action is thus $N_f=2$.
We used the following methods to optimise the performance of the 
HMC algorithm. Firstly the effective coupling $\beta^{\prime}$
used during integration of the equations of motion along a
microcanonical trajectory was tuned so as to maximise the acceptance rate of the
Monte Carlo procedure for a fixed microcanonical time-step $d\tau$. 
As the lattice size
was increased $d\tau$ had to be taken smaller and the optimal
$\beta^{\prime}$ approached $\beta$.
For example for the $N_f=2$ case on a $16^3$ lattice and coupling $\beta=0.60$
the choices $d\tau=0.025$ and
$\beta^{\prime}=0.6007$ gave acceptance rate greater than $90\%$ for all
bare masses of interest. To maintain this acceptance rate on a $32^3$ lattice
we used $d\tau=0.015$ and $\beta^{\prime}=0.6004$.
Secondly the Monte Carlo procedure was optimised by
choosing the trajectory length at random from a Poisson distribution
with mean $\bar\tau=1$. This method of optimisation, which guarantees
ergodicity, also decreases autocorrelation times dramatically.
Statistical errors in our measurements were 
calculated by jackknife
blocking, which accounts for autocorrelations in a raw dataset.
We found the errors remained stable as the number
of blocks was varied
from 10 to 50.

\section{Simulation Results}
\subsection{Chiral Symmetry Breaking}

First let us make a statement of the problem we are addressing.
Since a continuous symmetry is never broken 
spontaneously
on a finite volume, we are required to work with $m_0\not=0$ and study the
limiting behaviour as the explicit symmetry breaking is removed.
Since for an asymptotically-free theory
the UV behaviour is governed by the 
gaussian fixed point at the origin, then the continuum limit of the model lies 
in the limit $\beta\to\infty$, and all physical quantities should be expressible
in terms of the scale set by the dimensionful coupling $g$. To compare
simulation results taken at different couplings (lattice spacings), therefore,
it is natural to work in terms of dimensionless variables such as $\beta m_0$,
$L/\beta$, or
$\beta^2\langle\bar\Psi\Psi\rangle$. As the continuum limit is approached, 
data taken at different $\beta$ should collapse onto a single curve when plotted
in dimensionless units. 
Formally, 
chiral symmetry in QED$_3$ is broken if
\begin{equation}
\lim_{\beta\to\infty}\lim_{\beta m_0\to0}\lim_{L/\beta\to\infty}
\beta^2\langle\bar\Psi\Psi\rangle\not=0.
\end{equation}

In practice taking the limits in the required order is difficult, for several
reasons:
{\em(i)} The lattice itself distorts continuum physics
considerably unless the lattice spacing $a$ can be chosen small compared
to the relevant physical wavelengths in the system - as we already mentioned
in the previous section a non-zero lattice spacing breaks explicitly 
the continuum U$(2N_f)$ symmetry of the action;
{\em(ii)} the size of the lattice $L^3$ must be large not just in dimensionless
units but also compared to any 
dynamically generated correlations in the system; and 
{\em(iii)} the chiral extrapolation $m_0\to0$ requires some theoretical
prejudice.
If the volume is large enough, $m_0$ lies in the linear regime of
$\langle \bar\Psi\Psi(m_0) \rangle = \langle \bar\Psi\Psi(0) \rangle + m_0 
\langle \bar\Psi\Psi(0)\rangle^\prime +\ldots$ and linear extrapolations of 
the data work well. However, the chiral extrapolation is based on 
assumptions which are not necessarily valid when $m_0=0$. New physics
at $m_0=0$ may alter the situation. In our study despite the fact 
that in all our simulations we reached the ``linear regime'' we take
a conservative stance by defining 
an upper bound for the condensate which is half its value at the smallest 
$m_0$.

\begin{figure}[p]

                \centerline{ \epsfysize=3.2in
                             \epsfbox{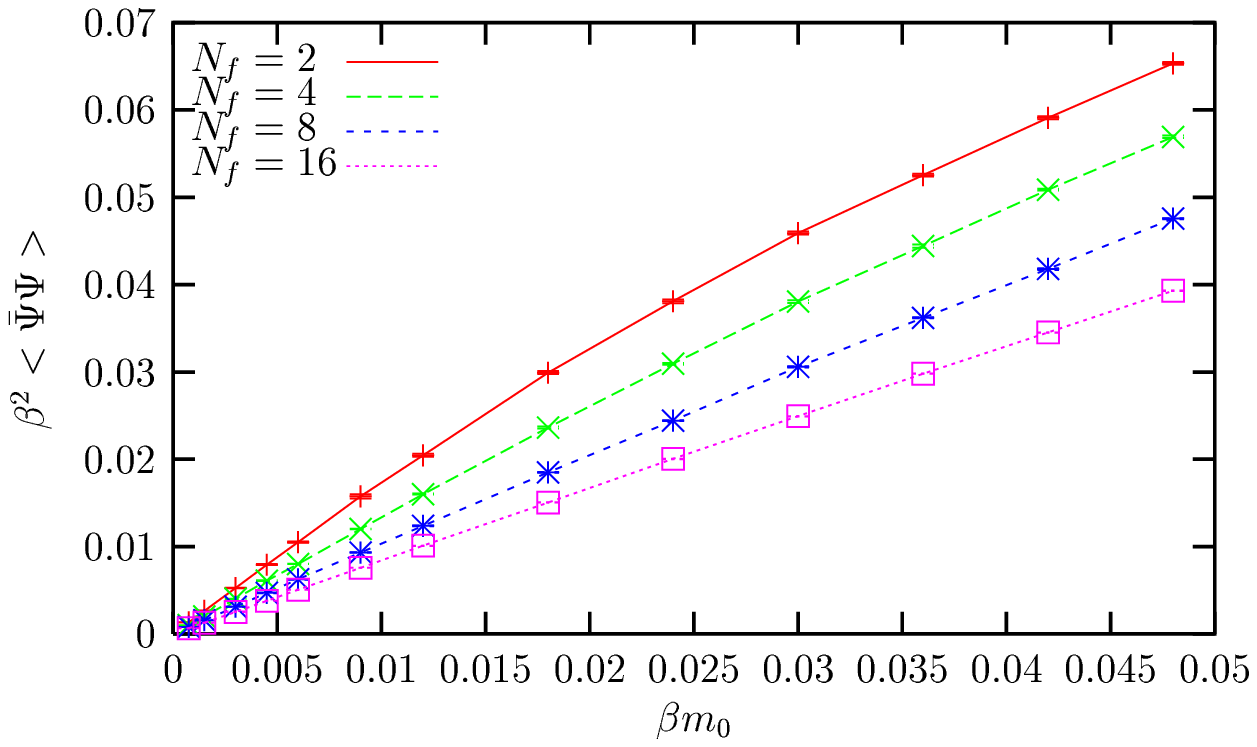}}

\smallskip
\caption[]{Dimensionless condensate $\beta^2 \langle \bar\Psi\Psi \rangle$
vs. dimensionless bare mass $\beta m_0$ for $N_f=2,4,8,16$, $\beta=0.6$ 
on a $16^3$ lattice.}
\label{fig:all_Nf}
\end{figure}

In Fig.~\ref{fig:all_Nf} we plot the
dimensionless chiral condensate $\beta^2 \langle \bar\Psi\Psi \rangle$ vs. the dimensionless bare mass
$\beta m_0$ for $N_f=2,4,8,16$.
The coupling $\beta=0.6$ and the lattice volume is $16^3$.
We generated approximately 1000 trajectories for each data point --
statistical error bars are generally smaller than the size of the symbols.
As $N_f$ increases the chiral condensate decreases (for $m_0 \geq 0$) because
of increased screening of the interaction by dynamical fermions.
As $m_0 \rightarrow 0$
all the curves tend to pass smoothly through the origin.
There is no sign of any discontinuous behaviour as $N_f$ is altered.
This motivates us to study in more detail the pattern of chiral symmetry 
breaking at small $N_f$ on larger volumes near the chiral limit.

\begin{figure}[p]

                \centerline{ \epsfysize=3.2in
                             \epsfbox{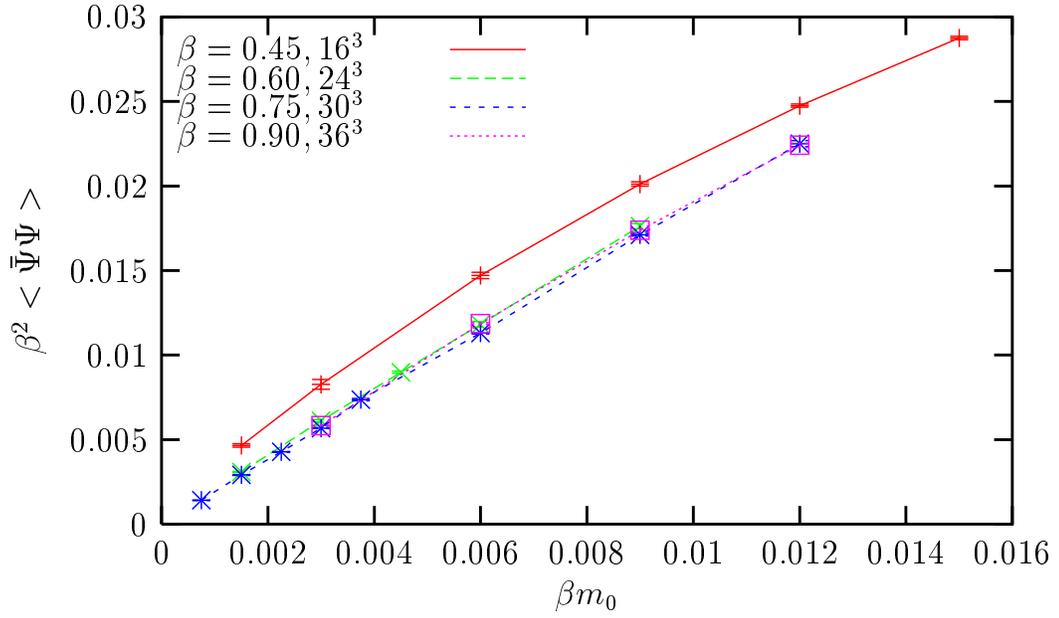}}

\smallskip
\caption[]{ $\beta^2 \langle \bar\Psi\Psi \rangle$ vs. $\beta m_0$ 
for $N_f=2$ at different
values of the coupling $\beta$ and constant physical volume $(L/\beta)^3$.}
\label{fig:discret}
\end{figure}

To check whether lattice data are characteristic of the continuum limit we 
next plot 
in Fig.~\ref{fig:discret} the dimensionless chiral condensate 
$\beta^2 \langle \bar\Psi\Psi \rangle$
vs. the dimensionless  fermion bare 
mass $\beta m_0$ for $N_f=2$ at different values of the coupling
$\beta=0.45,0.60,0.75,0.90$. 
In order to disentangle the lattice discretisation effects from the finite size effects we keep
the volume in physical units $(L/\beta)^3$ constant.
It can be easily inferred from the graph that at strong coupling ($\beta=0.45$) 
discretisation  effects are significant 
whereas for $\beta \geq 0.60$ the lattice 
artifacts become small since the data almost fall on the same line within 
the resolution of our analysis. 

\begin{figure}[p]

                \centerline{ \epsfysize=3.2in
                             \epsfbox{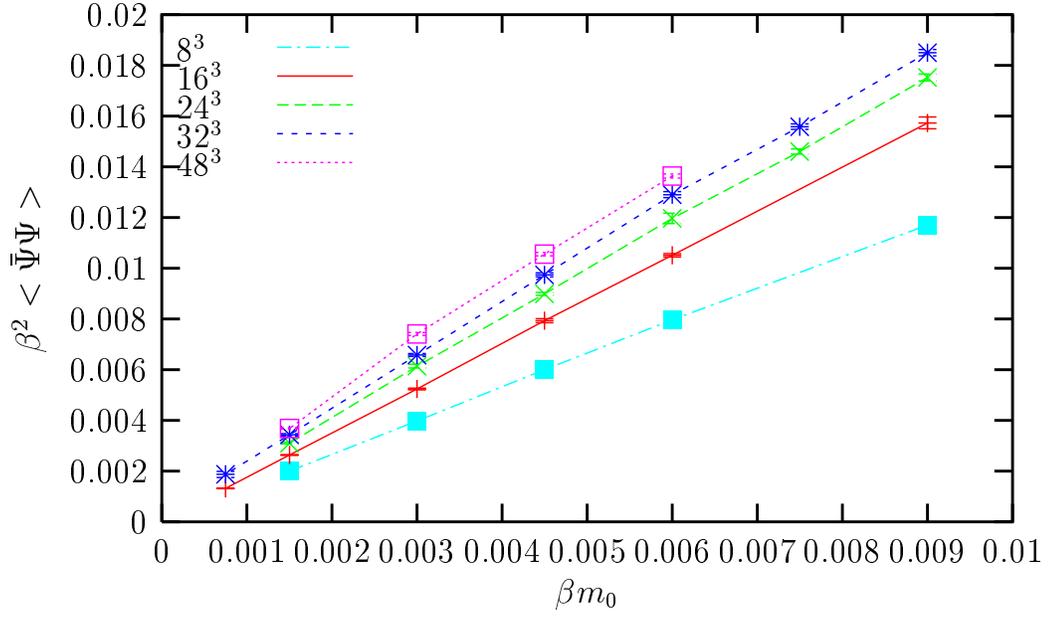}}

\smallskip
\caption[]{$\beta^2 \langle \bar\Psi\Psi \rangle$ vs. $\beta m_0$
 for $N_f=2$, $\beta=0.6$
and lattice sizes $8^3, 16^3, 24^3, 32^3, 48^3$.}
\label{fig:cond_b=0.6}
\vskip 5 truecm
\end{figure}

Having clarified the issue of the continuum limit we next investigate the 
thermodynamic limit, simulating with $N_f=2$ at $\beta=0.6$.
In Fig.~\ref{fig:cond_b=0.6} we present our results for 
$\beta^2\langle\bar\Psi\Psi\rangle$ vs. $\beta m_0$ 
on different lattice sizes: $8^3, 16^3, 24^3, 32^3$ and $48^3$. 
Note that the volumes and masses explored here are considerably
closer to the thermodynamic and chiral limits than those of the study of
\cite{alex}, for which $L/\beta\leq32$ and $\beta m_0\geq0.005$.
We generated $300 - 800$ configurations for each data point.
This enables us to expose finite size effects at different values of the 
fermion bare mass. 
Finite size effects 
become small (but still significant) for $L \geq 24$ ($L/\beta\geq40$),
which is consistent with the quenched results of \cite{hands};
however as $L$ increases all the lines continue to pass smoothly through
the origin, suggesting that chiral symmetry remains unbroken.

\begin{figure}[t]

                \centerline{ \epsfysize=3.2in
                             \epsfbox{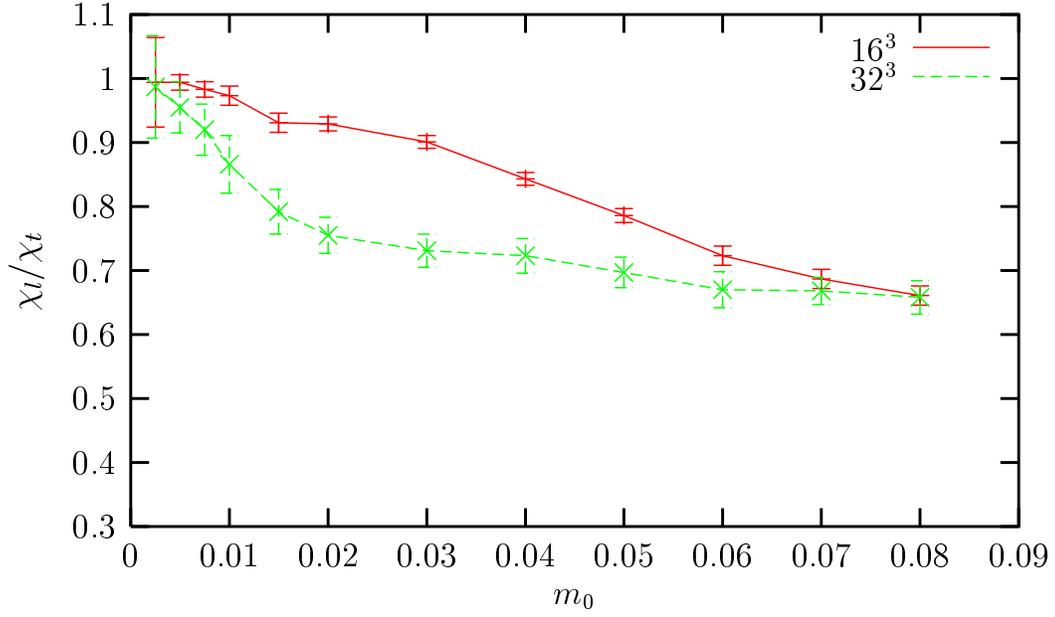}}

\smallskip
\caption[]{ The ratio of longitudinal to transverse susceptibilities
$R$ vs. $m_0$ for $N_f=2$, $\beta=0.6$ on $16^3$ and $32^3$ lattices.}
\label{fig:susc_b=0.6}
\end{figure}

\begin{figure}[b]

                \centerline{ \epsfysize=3.2in
                             \epsfbox{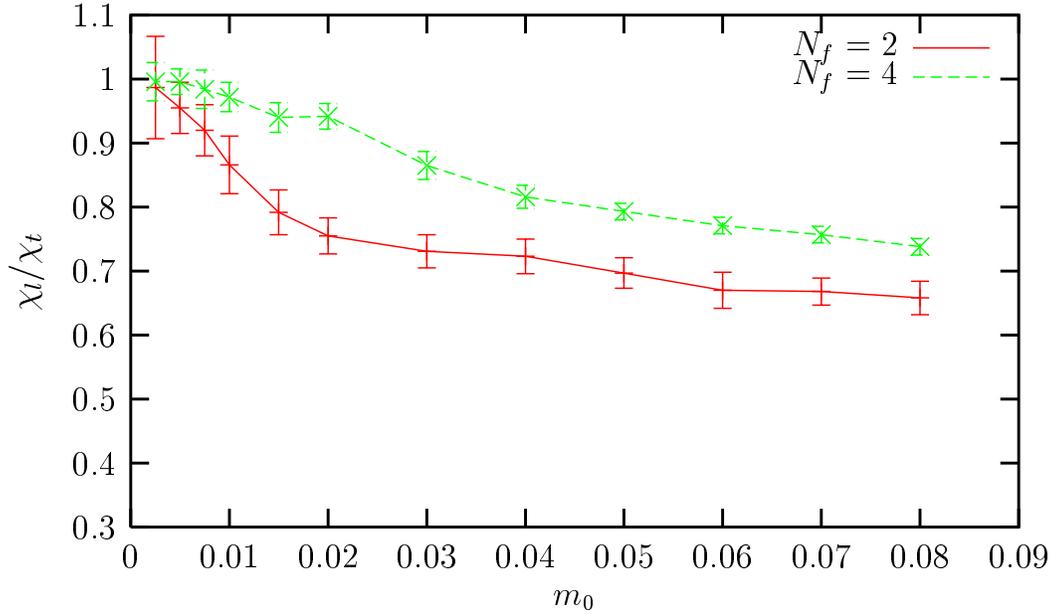}}

\smallskip
\caption[]{$R$ vs. $m_0$ for $N_f=2,4$, $\beta=0.6$ on $32^3$ lattices.}
\label{fig:susc.N=2.N=4}
\end{figure}

As well as the chiral condensate, we have also studied mesonic correlation
functions. In many cases these quantities yield useful information about the
nature of the ground state model, although they are also more prone to finite
volume effects than the simple order parameter.
Fig.~\ref{fig:susc_b=0.6} shows results from $N_f=2$, $\beta=0.6$ 
on lattice size $16^3$ and $32^3$
for the ratio
of longitudunal to transverse susceptibilities $R \equiv \chi_l/\chi_t$,
where $\chi_{l,t}$ are 
the integrated propagators in scalar and pseudoscalar
meson channels respectively:
\begin{equation}
\chi_l =  \sum_x \langle\bar\chi\chi(0)\bar\chi\chi(x)\rangle; \ \ \
\chi_t =  \sum_x\langle\bar\chi\varepsilon\chi(0)
\bar\chi\varepsilon\chi(x)\rangle.
\end{equation}
The longitudinal susceptibility $\chi_l$ has contributions
from diagrams with both connected and disconnected fermion lines;
we checked that the connected contribution is dominant
throughout our parameter space.
These quantities are much noisier than the
chiral condensate.
The transverse susceptibility is most conveniently estimated via the 
Ward identity $\chi_t = \langle \bar\Psi \Psi \rangle /m_0$.
In the chiral limit we expect
\begin{equation}
\lim_{m_0\to0}R=\left\{\begin{array}{ll}
  0, & \mbox{chiral symmetry broken;} \\
  1, & \mbox{chiral symmetry unbroken.}
  \end{array}
                \right. 
\end{equation}
As we can see from the figure, at large values of $m_0$
the finite size effects are small, but become 
significantly larger at intermediate values of $m_0$. 
This implies that finite volume effects are not fully under control;
the fact that $R$ from both $16^3$
and $32^3$ lattices appears to converge to 1 
as $m_0 \rightarrow 0$ decreases does not
necessarily imply chiral symmetry is unbroken.
It could be attributed to finite volume effects. Nonetheless, there is no
indication of any range of $m_0$ over which $dR/dm_0$ is positive,
required if
chiral symmetry is broken.
In Fig.~\ref{fig:susc.N=2.N=4} we plot $R$ vs. $m_0$ for $N_f=2,4$.
This result is compatible
with the data presented in Fig.~\ref{fig:all_Nf}, i.e. that 
the explicit chiral symmetry breaking at nonzero $m_0$, indicated by the
departure of $R$ from unity, is stronger as
$N_f$ decreases.

\begin{figure}[p]

                \centerline{ \epsfysize=3.2in
                             \epsfbox{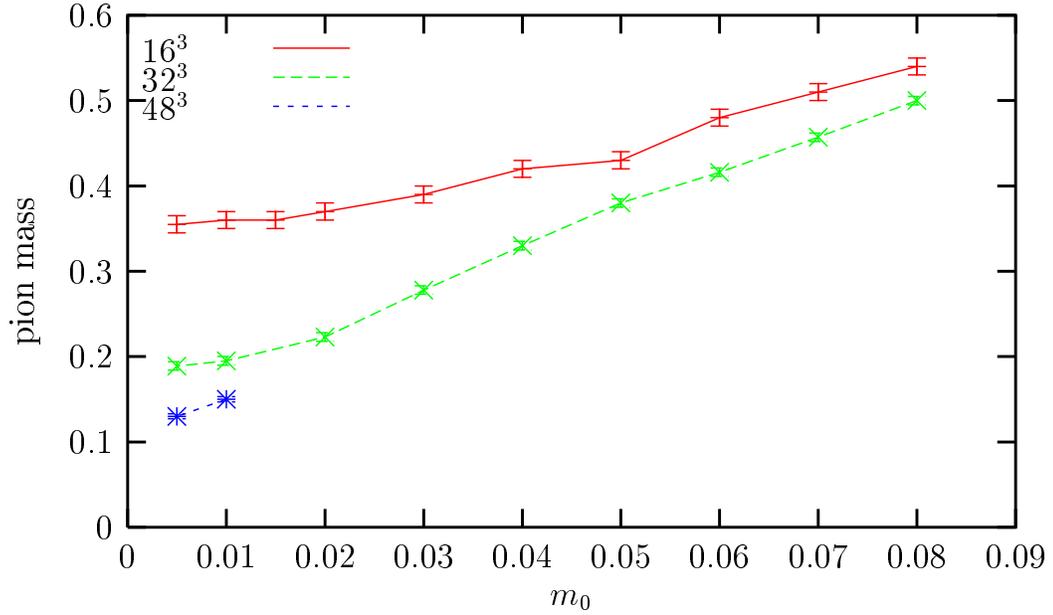}}

\smallskip
\caption[]{Pion mass vs. $m_0$ for $N_f=2$, $\beta=0.6$ on 
lattice sizes $16^3$, $32^3$ and $48^3$.}
\label{fig:pion}
\end{figure}

\begin{figure}[p]

                \centerline{ \epsfysize=3.2in
                             \epsfbox{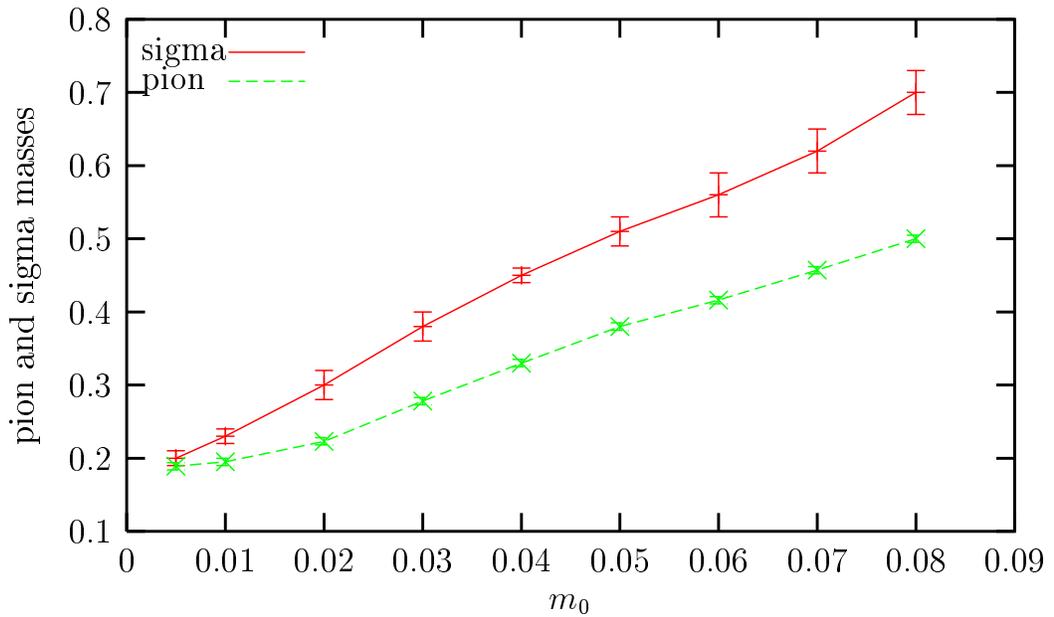}}

\smallskip
\caption[]{Masses of pion and sigma mesons vs. $m_0$ for $N_f=2$, $\beta=0.6$
on a $32^3$ lattice.}
\label{fig:pion_sigma}
\end{figure}

In Fig.~\ref{fig:pion} we present the results 
for the pion mass $M_{\pi}$ vs. fermion bare mass $m_0$ for
$N_f=2$, $\beta=0.6$ and lattice sizes $16^3, 32^3, 48^3$.
If the model is chirally symmetric we expect $M_{\pi} \propto m_0$;
otherwise from conventional PCAC arguments
we expect $M_{\pi} \propto \sqrt{m_0}$.
As in the case of $R$, $M_\pi$ suffers from
very strong finite volume effects and therefore 
decisive conclusions cannot be drawn.
In Fig.\ref{fig:pion_sigma} we plot the pion and sigma masses vs. $m_0$
for $N_f=2$, $\beta=0.6$ and lattice size $32^3$.
The masses tend to become equal as $m_0 \rightarrow 0$, and show no sign of
tending to zero in the chiral limit. It should be noted in this regard that 
there are SD calculations predicting the absence of light scalar excitations
for $N_f>N_{fc}$ \cite{ATW}.
However, as already stressed
no definitive conclusions can be drawn until
finite size effects are
under control. 

\begin{figure}[p]

                \centerline{ \epsfysize=3.2in
                             \epsfbox{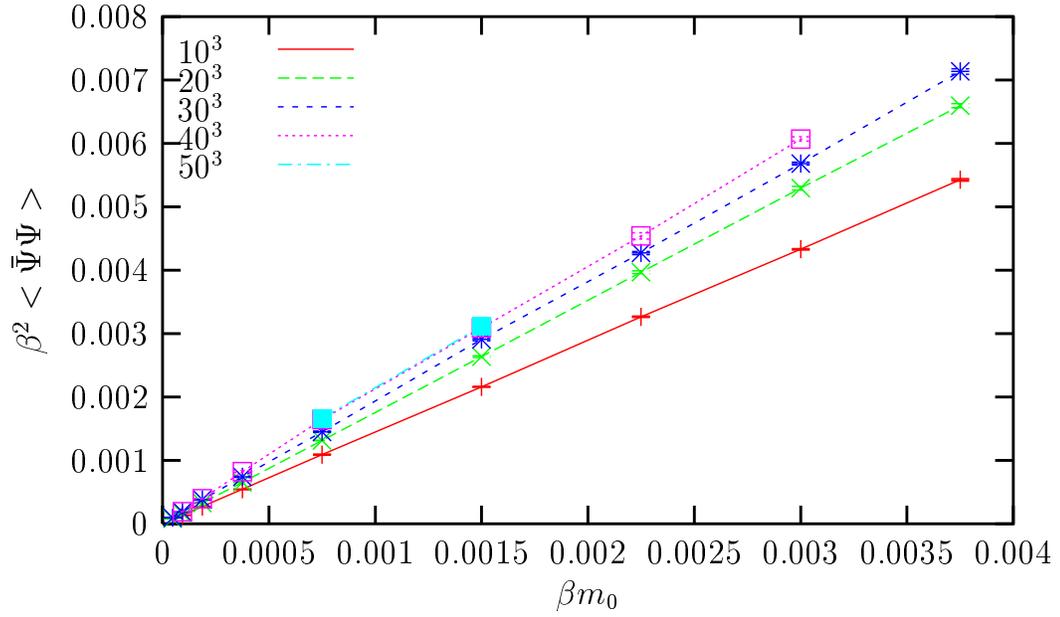}}

\smallskip
\caption[]{Condensate vs. bare mass for $N_f=2$, $\beta=0.75$ and lattice sizes $10^3,
20^3, 30^3, 40^3$ and $50^3$.}
\label{fig:cond_b=0.75}
\end{figure}

\begin{figure}[p]

                \centerline{ \epsfysize=3.2in
                            \epsfbox{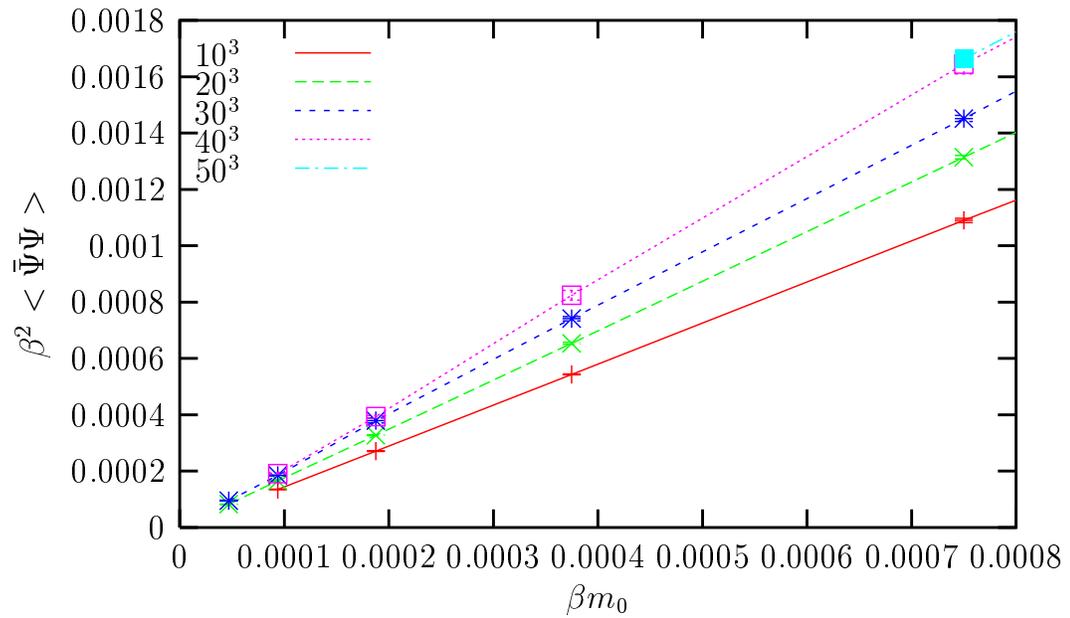}}
\smallskip
\caption[]{Same as Fig.\ref{fig:cond_b=0.75} but zoomed near the origin.}
\label{fig:cond_zoom}
\end{figure}

Since our best chance of controlling the extrapolation to infinite volume
appears to come from measurements of the order parameter, where signals 
are less noisy and finite volume effects relatively small, we now return to 
this and
discuss results from $N_f=2$ simulations at $\beta=0.75$, closer to the
continuum limit than our previous simulations.
These datasets result from between 300 and 500 HMC trajectories.
We studied the behaviour of the chiral 
condensate in detail in the large volume and small $m_0$ limits. 
Fig.~\ref{fig:cond_b=0.75} shows results for $\beta^2\langle\bar\Psi\Psi\rangle$
vs. $\beta m_0$
extracted from simulations on system sizes $10^3, 20^3, 30^3, 40^3$ and $50^3$.
These simulations were performed very close to the chiral limit
$(m_0 \leq 0.005)$. 
Although we don't extrapolate our data to the
chiral limit
it is clear that all the data tend to pass smoothly through the origin.
A comparison of the $40^3$ and $50^3$ points shows that finite volume effects
are under control, at least for $\beta m_0\geq0.0007$.
Fig.~\ref{fig:cond_zoom} shows the same data zoomed to the vicinity of
the origin. 
We can see that even when $m_0$ is very small, 
the data tend to pass smoothly through the origin: there is no indication of 
any tendency for the extrapolated intercept to be non-zero signalling
chiral symmetry breaking. With the conservative criterion adopted above
we conclude that for $N_f=2$, 
$\beta^2 \langle \bar \Psi \Psi \rangle \leq 5 \times 10^{-5}$, 
which is a strong indication 
that chiral symmetry in QED$_3$ may be restored for $N_f \geq 2$.

In Figs.~\ref{fig:cond_b=0.75} and \ref{fig:cond_zoom}
we plotted the condensate vs. 
bare mass keeping the lattice volume $L^3$ constant along each line. 
Hitherto we have assumed that the physical volume of the system is
proportional to $(L/\beta)^3$.
If dynamical mass generation occurs by some means, however,
so that a correlation length
$\xi<\infty$ develops,
then the correct measure of the physical volume is the ratio
$(L/\xi)^3$ (this is merely the statement that finite volume effects are
sensitive to which phase the theory resides in). 
This quantity varies along the 
constant lattice volume lines of
Figs.~\ref{fig:cond_b=0.75},\ref{fig:cond_zoom} because $\xi$ 
is in principle a function 
of both $m_0$ and $\beta$. 
We have attempted to study finite volume effects taking this possibility into
account:
since however we don't have an accurate estimate of a correlation length $\xi$
either from the sigma mass, because of strong finite volume effects, or from the
fermion mass, because the corresponding propagator is not gauge invariant, we
are forced to estimate the behaviour of $\xi(\beta,m_0)$ directly from the 
condensate data. Since the precise
relation between $\xi$ and $\langle\bar\Psi\Psi\rangle$ is unknown, 
we proceed by observing that in all of 
our datasets for fixed $\beta$ and $m_0\geq0.001$ it is approximately true that 
$\langle\bar\Psi\Psi\rangle\propto m_0$ and assuming that
$\xi\propto\langle\bar\Psi\Psi\rangle^{-\alpha}\propto m_0^{-\alpha}$ 
for some power
$\alpha=(d-1-\eta_{\bar\Psi\Psi})^{-1}$ where $\eta_{\bar\Psi\Psi}$ is the
anomalous scaling dimension of the composite operator $\bar\Psi\Psi$. 
We then cover the range of possibilities by evaluating
the condensate keeping $Lm_0^\alpha$
constant, with the choices $\alpha={1\over2}$ expected 
for a QCD-like theory in which $\eta_{\bar\Psi\Psi}$ is perturbatively small;
and $\alpha=1$,
expected for a theory such as the NJL model in which 
$\eta_{\bar\Psi\Psi}\approx d-2$ and 
the condensate
is directly proportional to the dynamically-generated scale.
A value
predicted for QED$_3$ with $N_f<N_{fc}$ 
by the SD approach is $\alpha={2\over3}$ corresponding
to $\eta_{\bar\Psi\Psi}={1\over2}$
\cite{Nfc}.

\begin{figure}[p]

                \centerline{ \epsfysize=3.2in
                             \epsfbox{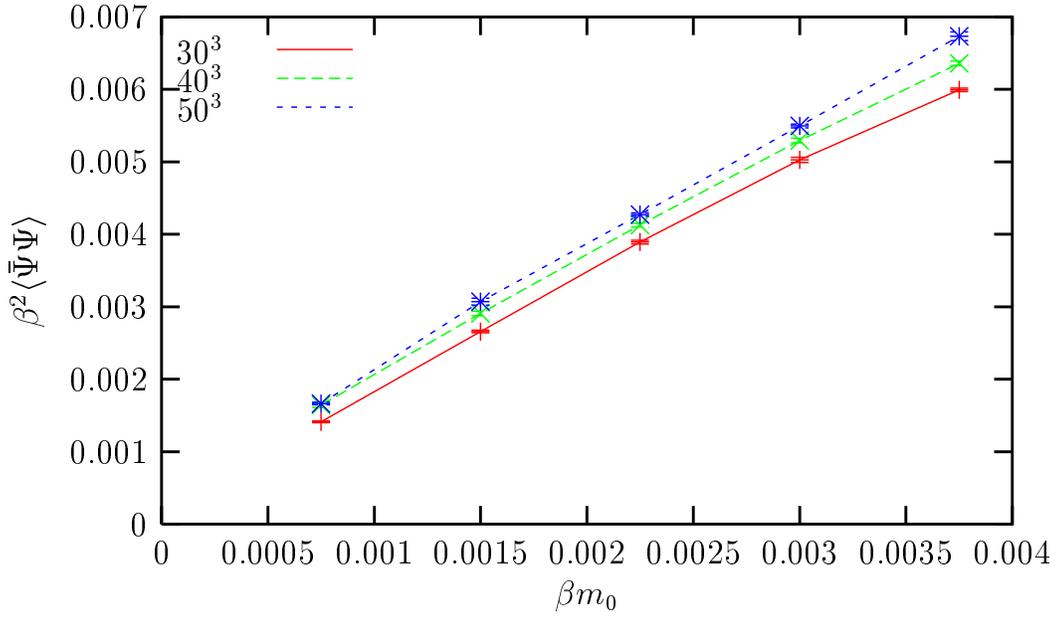}}

\smallskip
\caption[]{Condensate vs. fermion bare mass for $N_f=2$ and $\beta=0.75$ with
$L\sqrt{m_0}$ kept approximately constant along each curve.
The points 
at the smallest value of $\beta m_0$ were extracted from                 
simulations on $50^3, 40^3$ and $30^3$ lattices for each of the three different curves.}
\label{fig:norm_sqrt.m}
\end{figure}

\begin{figure}[p]

                \centerline{ \epsfysize=3.2in
                            \epsfbox{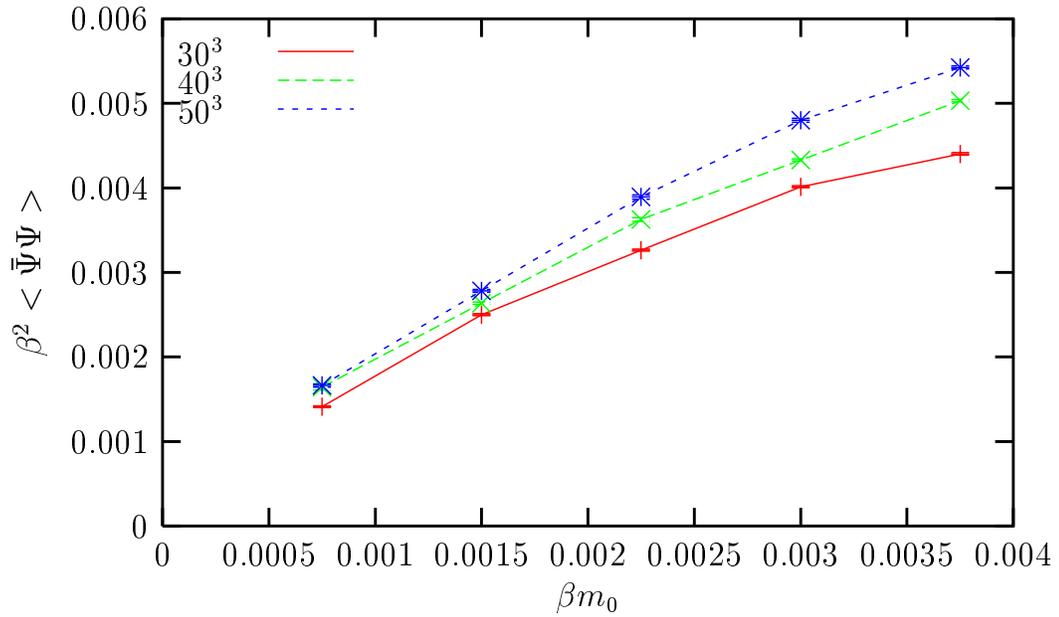}}
\smallskip
\caption[]{Condensate vs. fermion bare mass for $N_f=2$ and $\beta=0.75$ with
$Lm_0$ kept approximately constant along each curve. 
}
\label{fig:norm_m}
\end{figure}

The results for $\beta=0.75$ and $m_0 \geq 0$
are shown in Figs.~\ref{fig:norm_sqrt.m}
and \ref{fig:norm_m} respectively. In the case of the constant
$L\sqrt{m_0}$  normalisation the lattice sizes for the three curves shown are
$[50^3,36^3,30^3,26^3,22^3]$, $[40^3,28^3,24^3,20^3,18^3]$ and $[30^3
,22^3,18^3,16^3,14^3]$. For constant $Lm_0$ the 
sequences are
$[50^3,26^3,18^3,14^3,10^3]$, $[40^3
,20^3,14^3,10^3,8^3]$
and $[30^3,16^3,10^3,8^3,6^3]$.
By the very nature of the way the curves are produced an extrapolation 
to a non-trivial chiral limit is impossible, since eventually the assumption
that $\langle\bar\Psi\Psi\rangle\propto m_0$ must break down;
we note, however, that both sets of curves have 
negative curvature. It is also apparent that the curves
with constant $L\sqrt{m_0}$ appear to approach the infinite volume limit
uniformly over a wide range of $m_0$, whereas those keeping $Lm_0$ constant
diverge as $m_0$ increases. This is tentative evidence against a large value 
for $\eta_{\bar\Psi\Psi}$.

\subsection{Deconfinement of Fractional Charge}

Finally we consider a different issue, namely the confining properties
of the theory. Since the gauge degrees of freedom are non-compact, matter 
fields and/or test charges may in principle be 
defined with any value of electric
charge $\tilde g$ which need not be commensurate
with the dynamical fermion charge $g$. If the latter is the case, then
such fractional charges should be confined by the Coulomb potential
independent of the phase of the model, since long-range forces cannot in this
case be screened by virtual $f\bar f$ pairs. It has been pointed out, however,
that when the chiral symmetry of QED$_3$ is broken, either explicitly or
spontaneously, then there may be a transition to a phase where fractional charge
is deconfined at non-zero
temperature \cite{grignani}.

In brief, on a system with finite extent $T^{-1}$ in the Euclidean time
direction, the action (\ref{eq:contmodel}) has a global $Z$ symmetry under
non-periodic gauge transformations; in a static gauge this corresponds to 
invariance under 
\begin{equation}
A_0(\vec x)\mapsto A_0(\vec x) + {2\pi}{{nT}\over
g},\;\;\;n\in\mbox{Z}.\label{eq:Z}
\end{equation}
It is readily seen that worldlines of all integer-charged matter fields are
invariant under (\ref{eq:Z}). For fractionally-charged particles, however, this
is no longer the case.
The analysis of \cite{grignani} derives an effective
potential ${\cal V}(A_0)$ 
which is periodic under (\ref{eq:Z}); as $T$ increases
the barrier between non-equivalent minima of ${\cal V}$ grows until the system 
becomes trapped near one such minimum, breaking the $Z$ symmetry,
with the fractionally-charged Polyakov line
$\Pi_{\tilde g}(\vec x)=\exp i\tilde g\int_0^{T^{-1}} A_0(\vec x)dx_0$
acting as a gauge invariant order parameter. 
By analogy with Yang-Mills theory the symmetry
broken phase is identified as one in which fractional charge is deconfined.

\begin{figure}[p]

                \centerline{ \epsfysize=4.2in
                            \epsfbox{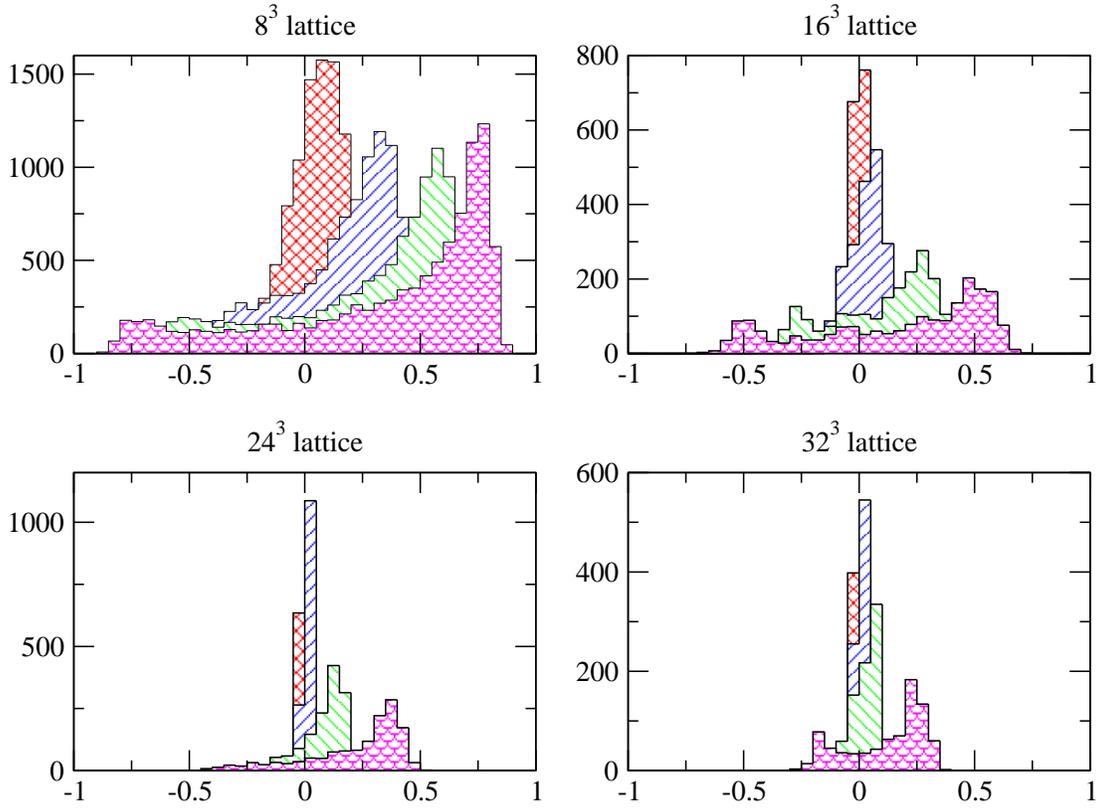}}
\smallskip
\caption[]{Histograms of the distribution of $\mbox{Re}(\Pi_n)^n$ from
simulations with $N_f=2$, $\beta=0.6$ on $8^3$, 
$16^3$, $24^3$ and $32^3$ lattices, from respectively 10000, 2000, 1400 and 800
configurations. The squared shading denotes $n=1$, cross-hatched to
north-east $n=2$, to north-west $n=4$, and scalloped $n=8$.
}
\label{fig:fourpolys}
\end{figure}

In the simulations of this paper we have not examined systems with the Euclidean
time direction distinguished; however, we have found it interesting to check
whether the $Z$ symmetry is broken as a means of monitoring finite volume effects.
We have chosen to examine charges which are rational fractions of the
fundamental charge: $\tilde g=g/n$, and define the
Polyakov loop $\Pi_n(\vec{x})$ by
\begin{equation}
\Pi_n(\vec x)=\prod_{t=1}^{L_t}\exp\biggl({i\over n}\theta_0(\vec x,t)\biggr).
\end{equation}
The original $Z$ symmetry (\ref{eq:Z}) translates into a $Z_n$ symmetry on 
$\Pi_n$; spontaneous breaking is therefore signalled by a non-zero expectation
of $\mbox{Re}(L^{-3}\sum_{\vec x}\Pi_n(\vec x))^n$. 
In Fig.~\ref{fig:fourpolys} we plot distributions of this quantity for
$n=1,2,4$ and 8 on a sequence of lattice volumes. On all systems the $n=1$ data
is sharply peaked about zero, indicating that there is no 
deconfinement of the fundamental integer charge\footnote{More precisely,
it suggests the free energy of an isolated fundamental charge diverges in the
thermodynamic limit, which could also occur in the conformal phase.}. 
On $8^3$, however, the
distributions for all $n\geq2$ are skewed in the positive direction, indicating
deconfinement. As the volume is increased, permitting fluctuations of the fields
to develop over longer distances, the deconfinement signal is washed out until
on $32^3$ only charges with $\tilde g=g/8$ remain deconfined. A possible
explanation of why smaller charges are harder to confine in a finite volume
is that $\tilde
g^2$
itself provides the scale in the logarithmic potential.

\section{Summary and Outlook}
In our study of QED$_3$ with $N_f \geq 2$
we attempted to establish whether chiral symmetry is broken or not
by studying various observables close to the continuum limit
$g \rightarrow 0$, on different  volumes in order to detect and control
finite size effects and near the chiral limit
$m_0 \rightarrow 0$. As expected, susceptibilities and meson masses suffer from
strong finite size effects
and by themselves they do not allow us to reach a definitive conculsion.
However, for the chiral condensate the continuum, thermodynamic 
and chiral limits are under better control. On the volumes we have been able to
study we have seen no evidence for
chiral symmetry breaking for any $N_f\geq2$.

Our upper bound for the condensate in the $N_f=2$ case is
$\beta^2\langle \bar \Psi \Psi \rangle \leq 5 \times 10^{-5}$, to be compared
with the estimate for the quenched model
$\beta^2\langle\bar\Psi\Psi\rangle\approx5\times10^{-3}$ \cite{sasha,hands}.
In addition, for all $N_f\geq2$ the curves of 
$\beta^2 \langle \bar \Psi \Psi \rangle$ vs.
$\beta m_0$ tend to pass smoothly through the origin with no sign of any
discontinuous behaviour in $N_f$. This is evidence against the suggestion that 
the IR fixed point behaviour of QED$_3$ for $N_f\leq N_{fc}$
coincides with the UV behaviour of the 
three dimensional Thirring model as suggested by both the large-$N_f$ expansion
and SD studies \cite{deldebbio}. For the Thirring model
the critical $N_{fc}$ below which 
a non-trivial UV fixed point exists has been estimated by Monte Carlo simulation 
to be $N_{fc}\simeq5$ \cite{deldebbio,lucini}. The two models appear to lie in
different universality classes, possibly because for $N_f<N_{fc}$
the Thirring model has no
massless degree of freedom corresponding to the photon.

We are currently simulating the model with $N_f=1$ to check whether in 
this case chiral symmetry is broken, as predicted by \cite{appelquist}. 
It will also be interesting in this respect 
to study whether the
continuum U$(2N_f)$ symmetry is restored; this may entail the 
use of improved fermion actions \cite{MILC} in future simulations.
We also plan to extend our work to study 
in detail the deconfinement phase transition of fractional electric 
charges at $T>0$.

\newpage
\section*{Acknowledgements}
SJH and CGS were supported by a Leverhulme Trust grant, and 
JBK in part by NSF grant PHY-0102409. 
The computer simulations were done on the Cray SV1's at NERSC, the IBM-SP
at NPACI, 
and on the SGI Origin 2000 at the University of Wales Swansea.  
We have enjoyed discussions with Ian Aitchison, Nick Mavromatos,
Sarben Sarkar and Rohana Wijewardhana.

\newpage

\newpage

\newpage

\newpage

\newpage

\newpage

\newpage

\newpage

\end{document}